\documentclass[
 reprint,amsmath,amssymb,aps]{revtex4-1}

\usepackage{graphicx}
\usepackage{dcolumn}
\usepackage{bm}
\usepackage{epsfig}
\usepackage{subfigure}
\usepackage{multirow}
\usepackage{amsmath}

\begin{document}

\title{Symmetry based parametrizations of the lepton mixing matrix}

\author{Sanjeev Kumar}\email{skverma@physics.du.ac.in}
 \affiliation{Department of Physics and Astrophysics, University of Delhi,\\ 
Delhi 110007, India.} 
 
\date{\today}

\begin{abstract}
There are many mixing schemes based upon flavor symmetries that predict a \mbox{vanishing $\theta_{13}$}. These mixing schemes need corrections or modifications to 
account for recent experimental measurements of nonzero $\theta_{13}$. 
We propose new parametrizations for the lepton mixing matrix to quantify the minimal 
modifications needed in these mixing schemes. The parametrizations can be factorized in
two parts: $U_{0}(a,b)$ and $R(\theta,\phi)$. The first factor can be viewed
as a zeroth order mixing matrix coming from some flavor symmetry. 
It reproduces the popular mixing schemes 
based upon flavor symmetries for suitable values of $a$ and $b$. The second factor can 
be interpreted as a minimal modification to the mixing matrix and is responsible for nonzero $\theta_{13}$, nonmaximal $\theta_{23}$ and \textit{CP} violation. We also find the
experimentally allowed parameter space for the parameters $a$ and $b$ and compare
it with the symmetry based values for these parameters.
\end{abstract}


\maketitle

\section{Introduction} 

The vastly different fermion masses and mixings point to a broken flavor symmetry
that might be hidden in the family structure of fermions. In a class of lepton mass models, 
the residual flavor symmetry in the lepton mass matrices  can be related to the original 
flavor symmetry $\mathcal{G}$ of the \mbox{Lagrangian \cite{dynamics,origin}}. In such 
models, two different symmetries in the charged lepton and neutrino sector, 
$G_{l}$ and $G_{\nu}$, are preserved when the original symmetry  $\mathcal{G}$ is 
broken. These different symmetries are responsible for different diagonalization
matrices for the charge lepton mass matrix $M_{l}$ and effective neutrino mass
matrix $M_{\nu}$. Thus, a flavor symmetry can predict the neutrino mixing 
\mbox{ matrix $U$}.  As an illustration, bimaximal mixing \mbox{(BM) \cite{bm}} and 
tribimaximal mixing \mbox{(TBM) \cite {tbm}} 
are based upon the symmetry group $S_{4}$ or  a larger group containing 
$S_{4}$ as a subgroup \mbox{\cite{dynamics, s4-tbm}}. 

The mixing schemes like BM and TBM are called full mixing schemes since they predict all 
the three columns of \mbox{$U$ \cite{dynamics}}. 
Other examples are the golden ratio mixing of 
\mbox{type I (GRM1) \cite{grm1}} and \mbox{type II (GRM2) \cite{grm2}}, hexagonal 
mixing \mbox{(HM) \cite{hm}} and democratic mixing \mbox{(DM) \cite {dm}}.
These mixing schemes predict a vanishing $\theta_{13}$ and maximal $\theta_{23}$ and 
can at best serve as leading-order approximations to the neutrino mixing matrix.
The mass models producing these mixing schemes predict corrections to the mixing angles
at next to leading order that are usually of the same magnitude for all angles. 
This makes it difficult to accommodate a relatively large $\theta_{13}$ in the lepton 
mass models based upon complete mixing that predict zero $\theta_{13}$ at the leading 
order.

Another way of having deviations from a full mixing scheme was suggested by 
\mbox{Lam \cite{dynamics, mit}} in the form of partial mixing schemes. A partial mixing 
matrix of type $C_{i}$ ($R_{i}$) is defined as a unitary matrix with the $i$th column (row) 
fixed to $N\{a~b~1\}^T$ ($N\{1~b~a\}$), while keeping the other two columns free 
within unitarity 
constraints. Here, the parameters $a$ and $b$ are fixed by the flavor symmetry that is 
responsible for the corresponding complete mixing scheme and $N=1/\sqrt{1+a^{2}+b^{2}}$
is the normalization constant.
For example,  $\mu$-$\tau$ symmetry is a partial mixing of type $C_{3}$ with $a=0$
and $b=1$ and trimaximal \mbox{mixing (TM) \cite{tm}} 
is a partial mixing of type $C_{2}$ with 
$a=1$ and $b=1$. Similarly, one can obtain several partial mixing matrices of the types 
$C_{i}$ and $R_{i}$ from complete mixing schemes like  BM, TBM, DM, GRM1, GRM2, DM, 
and HM by the selecting respective values of $a$ and $b$ listed in \mbox{Table 1}.
We remark that the value of $\theta_{13}$ in the partial mixing schemes $C_{3}$ and 
$R_{1}$ remains unaltered from its value in the corresponding complete mixing scheme.
Hence, these schemes are not of much interest to our work and have not been tabulated in
\mbox{Table 1}.

The recent conclusive measurements of a finite \mbox{$\theta_{13}$ 
\cite{t2k, minos,daya-bay,reno}} have initiated an exploration of 
new leading order approximations to the mixing matrix with nonzero $\theta_{13}$ 
and nonmaximal $\theta_{23}$ that could result from some larger symmetry groups. 
Two mixing schemes based upon the modular group were proposed recently by Toorop, 
Feruglio and Hagedorn \mbox{\cite{modular}} which will be called the TFH1 and TFH2 mixing 
schemes. Again, these mixing schemes need next to order corrections to explain 
three mixing angles simultaneously. One way to do this is by constructing partial mixing 
schemes for these mixing patterns. We have also included these mixing schemes in \mbox{Table 1}.

The plan of this paper is as follows. In Sec. 2, we review the link between the residual 
symmetry of the neutrino mass matrix and the neutrino mixing matrix. In Sec. 3, 
we first present six parametrizations for the neutrino mixing matrix with four free 
parameters. These parametrizations are best suited to describe the partial mixing 
matrices and can be used to study the corrections in the popular mixing schemes mentioned 
above. There are many recent studies where similar modifications to various
mixing matrices have been \mbox{studied \cite{modifications}}. 
Here, we use our parametrizations and obtain sum rules for mixing angles and \textit{\textit{CP}} 
violation in a model-independent manner. In Sec. 4, we show that these 
parametrizations can be factorized in two parts: the first part 
can be identified with the complete mixing schemes and the second part can be 
thought of a minimal modification or perturbation to these mixing 
schemes. In Sec. 5, we perform a model-independent analysis for the allowed parameter 
space of these parametrizations in light of current experimental data and highlight 
the implications.  

\begin{table*}[t]
\begin{center}
\begin{tabular}{ccccccc}
\hline 
\hline
Mixing & &  \multicolumn{2}{c}{$C_{1}$ }  & & \multicolumn{2}{c}{$C_{2}$ } \\
\cline{3-4}\cline{6-7}
pattern && $a$ & $b$ && $a$ & $b$  \\
\cline{1-1}\cline{3-4}\cline{6-7} \\
BM && $\sqrt{2}$ & $1$  && $\sqrt{2}$ & $1$ \\
TBM && $2$ & $1$    && $1$ & $1$\\
DM && \small{$\sqrt{\frac{3}{2}}$ } & $\frac{1}{\sqrt{2}}$   && 
\small{$\sqrt{\frac{3}{2}}$} & $\frac{1}{\sqrt{2}}$  \\
GRM1 && \small{$\sqrt{3+\sqrt{5}}$ } & $1$  && \small{$\sqrt{3-\sqrt{5}}$} & $1$\\
GRM2 && \small{$\sqrt{2+\frac{4}{\sqrt{5}}}$} & $1$  && $\sqrt{10-4 \sqrt{5}}$ & $1$ 
\\
HM && $\sqrt{6}$ & $1$  && $\sqrt{\frac{2}{3}}$ & $1$ \\
TFH1 && $\frac{1}{2} \left(\sqrt{3}+1\right)$ & $\frac{1}{2} \left(\sqrt{3}-1\right)$  && 
$1$ & $1$ \\
TFH2 && $2+\sqrt{3}$ & $1+\sqrt{3}$  && $1$ & $1$ \\
\hline 
\hline
Mixing & &  \multicolumn{2}{c}{$R_{2}$ }  & & \multicolumn{2}{c}{$R_{3}$ } \\
\cline{3-4}\cline{6-7}
pattern && $a$ & $b$ && $a$ & $b$  \\
\cline{1-1}\cline{3-4}\cline{6-7} \\
BM && $\sqrt{2}$ & $1$  && $\sqrt{2}$ & $1$ \\
TBM && $\sqrt{2}$ & $\sqrt{2}$    && $\sqrt{3}$ & $\sqrt{2}$ \\
DM && $2$ & $1$   && $1$ & $1$ \\
GRM1 && \small{$\sqrt{\frac{1}{2} \left(5+\sqrt{5}\right)}$} & 
\small{$\sqrt{\frac{1}{2} \left(3+\sqrt{5}\right)} $} && 
\small{$\sqrt{\frac{1}{2} \left(5+\sqrt{5}\right)}$} & 
\small{$\sqrt{\frac{1}{2} \left(3+\sqrt{5}\right)}$ } \\
GRM2 && \small{$\sqrt{2+\frac{2}{\sqrt{5}}}$}
 & \small{$\frac{1+\sqrt{5}}{\sqrt{10-2 \sqrt{5}}}$ } && 
 \small{$\sqrt{2+\frac{2}{\sqrt{5}}}$}
 & \small{$\frac{1+\sqrt{5}}{\sqrt{10-2 \sqrt{5}}}$ } \\
HM && $1$ & $\sqrt{3}$  && $1$ & $\sqrt{3}$  \\
TFH1 && $2+\sqrt{3}$ & $1+\sqrt{3}$  && 
$1$ & $1$ \\
TFH2 && $1$ & $1$  && $2+\sqrt{3}$ & $1+\sqrt{3}$ \\
\hline
\hline
\end{tabular}
\end{center}
\caption{The values of the parameters $a$ and $b$ for the partial mixing schemes
matrices of type $C_{2}$, $C_{3}$, $R_{2}$ and $R_{3}$ generated from several
popular complete mixing schemes. The $j$th column (row) of the mixing matrix of type
$C_{j}$ ($R_{j}$) is given by $N\{a~b~1\}^T$ ($N\{1~b~a\}$) and the other two columns
(rows) are free within unitarity constraints.}
\end{table*}

\section{Flavor symmetry and lepton mixing}

If $\mathcal{G}_{l}$ is the residual symmetry of the 
charged lepton mass matrix $M_l$ and $\mathcal{G}_{\nu}$ is the residual 
symmetry of the effective neutrino mass matrix $M_{\nu}$,
$M_{l}$ and $M_{\nu}$ are invariant under the residual symmetry transformations: $F^{\dagger}M_lF=M_l$ and $G_{i}^TM_{\nu}G_{i}=M_{\nu}$  ($i=1,2$ and $3$). 
Here, the symmetry transformation $F$ is the 
diagonal generator of the group $\mathcal{G}_{l}=Z_n$ ($n\ge 3$) in the diagonal 
$M_l$ basis and the symmetry transformations $G_i$'s 
\begin{eqnarray}
G_1&=&U~diag(1,-1,-1)~U^{\dagger}  \nonumber \\
G_2&=&U~diag(-1,1,-1)~U^{\dagger} \\
G_3&=&U~diag(-1,-1,1)~U^{\dagger}, \nonumber 
\end{eqnarray}
form the group $\mathcal{G}_{\nu}=Z_2\times Z_2$ \cite{dynamics,origin}. 
The residual symmetries, described by the subgroups $\mathcal{G}_{l}$ and $\mathcal{G_{\nu}}$,  are remnants of the original flavor symmetry $\mathcal{G}$. 
When the $\mathcal{G}$ is broken into two 
different symmetry groups, the charged lepton mass matrix and the neutrino mass matrix 
may still be invariant under generators of the two residual groups 
$\mathcal{G}_{l}$ and $\mathcal{G_{\nu}}$. In a dynamic model, 
this is assured if the vacuum  alignments of the flavon fields are the invariant 
eigenvectors of the generators of the symmetry groups of the residual flavor 
\mbox{symmetry \cite{dynamics,origin}}. 
To obtain a full mixing matrix, the flavon fields coupling with neutrinos must satisfy three 
invariance conditions whereas to  obtain partial mixing matrix, they have to satisfy only 
one invariance condition. Hence, the partial mixing is less restrictive than full mixing and ideal for the present experimental scenario.

Several models exist in literature that predict partial mixing schemes of types
\mbox{$C_{1}$ \cite{tm1}} and \mbox{$C_{2}$ \cite {tm2}} obtained from TBM mixing. 
A simple recipe that transforms a TBM model to other mixing schemes like GR is given in
\mbox{Ref. \cite{recipe}}. A general approach based upon group theory for the construction of a dynamic model based upon type I and type II seesaw mechanisms is discussed in 
\mbox{Ref. \cite{dynamics}}. A generalization of $Z_{2}\times Z_{2}$ symmetry
in the neutrino mass matrix can give rise to partial mixings of type 
$C_{1}$ and \mbox{$C_{2}$ \cite{z2z2}}.

One advantage of such models is that they yield relations between the mixing angles and 
the \textit{\textit{CP}} violating phase instead of completely fixing the mixing angles. Such phenomenological 
relations have already been studied for special cases such as the trimaximal mixing  \mbox{matrix \cite{tm}}, matrices obtained from $Z_{2}$ \mbox{symmetry \cite{z2}}, 
and a partial mixing scheme obtained from the first column of the 
tribimaximal \mbox{mixing \cite{tm1}}. 
However, a general parametrization applicable for all such partial
mixing matrices does not exist at present. Consequently, the phenomenological relations resulting from partial mixing have also not been generalized so that they can be
studied in a model independent context. 
Such model-independent relations will be verifiable in future neutrino 
experiments  \cite{ t2k,nova} that are sensitive to the $\theta_{23}$ octant and \textit{CP} violation in neutrino oscillations. Hence, they can be used to distinguish different 
mixing schemes resulting from the different residual flavor symmetries and, if possible, to  reconstruct the original flavor symmetry. 

\section{The parametrizations}

The most general mixing matrix with the first column fixed to $N\{a~b~1\}^T$ can be written as
\begin{equation}
U_{(23)}=\left(
\begin{array}{ccc}
a N & N\sqrt{1+b^2} \cos \theta &
   N\sqrt{1+b^2} \sin \theta\\
 b N & \frac{e^{i \phi} \sin \theta-a b N\cos \theta}{\sqrt{1+b^2}} &
   -\frac{e^{i \phi} \cos \theta+a b N \sin
   \theta}{\sqrt{1+b^2}} \\
  N &-\frac{a N \cos \theta+b e^{i \phi} \sin \theta}{\sqrt{1+b^2}} &
   \frac{b e^{i \phi} \cos \theta-a N \sin
   \theta}{\sqrt{1+b^2}}
\end{array}
\right).
\end{equation}
For the mathematical proof, the reader is directed to the Appendix.
The other parametrization with the second column fixed to $N\{a~b~1\}^T$
is given by
\begin{equation}
U_{(13)}=\left(
\begin{array}{ccc}
 N\sqrt{1+b^2} \cos \theta &a N &
   N\sqrt{1+b^2} \sin \theta\\
 \frac{e^{i \phi} \sin \theta-a b N\cos \theta}{\sqrt{1+b^2}} &b N & 
   -\frac{e^{i \phi} \cos \theta+a b N \sin
   \theta}{\sqrt{1+b^2}} \\
-\frac{a N \cos \theta+b e^{i \phi} \sin \theta}{\sqrt{1+b^2}} &  N &
   \frac{b e^{i \phi} \cos \theta-a N \sin
   \theta}{\sqrt{1+b^2}}
\end{array}
\right).
\end{equation}
A nice property of these parametrizations is that the $(1,3)$ element vanishes
in the special case $\theta = 0$. 
We do not consider the parametrization $U_{(12)}$ because its third
column will be $N\{a~b~1\}^T$ and, therefore, the $(1,3)$ element does not vanish for 
$\theta = 0$.

Similarly, the parametrizations with the second
and third rows being equal to $N\{1~b~a\}$ are given by
\begin{widetext}
\begin{equation}
U^{(13)}=\left(
\begin{array}{ccc}
 \frac{b e^{i \phi} \cos \theta-a N \sin
   \theta}{\sqrt{1+b^2}} & 
   -\frac{e^{i \phi}\cos \theta+a b N \sin \theta}{\sqrt{1+b^2}}
   & N \sqrt{1+b^2} \sin \theta \\
 N & b N & a N \\
 -\frac{a N \cos \theta+b e^{i \phi} \sin
   \theta}{\sqrt{1+b^2}} & \frac{e^{i \phi}
   \sin \theta-a b N \cos \theta}{\sqrt{1+b^2}}
   & N \sqrt{1+b^2} \cos \theta
\end{array}
\right)
\end{equation}
and
\begin{equation}
U^{(12)}=\left(
\begin{array}{ccc}
 \frac{b e^{i \phi} \cos \theta-a N \sin
   \theta}{\sqrt{1+b^2}} & 
   -\frac{e^{i \phi}
   \cos \theta+a b N \sin \theta}{\sqrt{1+b^2}}
   & N\sqrt{1+b^2} \sin \theta \\
 -\frac{a N \cos \theta+b e^{i \phi} \sin
   \theta}{\sqrt{1+b^2}} & \frac{e^{i \phi}
   \sin \theta-a b N \cos \theta}{\sqrt{1+b^2}}
   & N \sqrt{1+b^2} \cos \theta \\
   N & b N & a N 
\end{array}
\right),
\end{equation}
\end{widetext}
respectively. Again, the $(1,3)$ element of the mixing matrix vanishes for 
$\theta = 0$ in these parametrizations. 
The parametrization $U^{(23)}$ with first row equal to $N\{1~b~a\}$
will not share this property and, hence, is not studied here. We further note that
many other parametrizations of the similar nature can be constructed by appropriate
permutations of the elements of the above mixing matrices. The choices for the positions of the elements we have made are not unique but motivated by simplicity of results.

If we substitute the values of $a$ and $b$ in the parametrizations $U_{(23)}$, 
$U_{(13)}$, $U^{(13)}$ and $U^{(12)}$ that are listed in \mbox{Table 1}, we will get the corresponding partial mixing schemes of types $C_{1}$, $C_{2}$, $R_{2}$ and $R_{3}$,
respectively. Further, if we put $\theta=0$, $\phi=0$ or $\pi$, and the values of 
$a$ and $b$ listed in Table 1, we get the corresponding complete mixing schemes. A partial 
mixing scheme, in our parametrizations, fixes $a$ and $b$ to the corresponding values 
listed in Table 1 while keeping $\theta$ and $\phi$ free. Therefore, the values of $a$ and 
$b$ listed in \mbox{Table 1} may be viewed as predictions of some flavor symmetries 
corresponding to respective complete mixing patterns. However, the parameters $a$ and 
$b$ are two of the parameters of the mixing matrix in our parametrization and are to be 
determined from the experimental data. The comparison between the experimentally 
allowed values of $a$ and $b$ with the symmetry based values given in \mbox{Table 1} 
form the bases of our model-independent analysis.

Just like all other unitary parametrizations for the mixing matrix, 
the above parametrizations have four free parameters $a$, $b$, $\theta$ and $\phi$. 
The mixing angles $\theta_{12}$, $\theta_{23}$ and $\theta_{13}$ and \textit{CP} 
violating phase $\delta$ in the Particle Data Group \cite{pdg} parametrization can be expressed in terms of these four parameters. 
The expressions for the mixing angle $\theta_{13}$ and the Jarlskog rephasing invariant 
measure of \textit{CP} violation, $J=Im(U_{11}U^*_{12}U^*_{21}U_{22})$ \cite{jcp},  
are identical for all the four parametrizations: 
\begin{equation}
\sin^2{\theta_{13}}=N^{2}\left(1+b^2\right) \sin ^2\theta
\end{equation}
and
\begin{equation}
J=B \sin\phi
\end{equation}
where
\begin{equation}
B=a b c N^{3}\sin \theta\cos\theta.
\end{equation}
The expressions for the other two mixing angles are different in different 
parametrizations and have been listed in \mbox{Table 2}. The parameters $\theta_{0}$
and $A$ used in \mbox{Table 2} are given by
\begin{equation}
\sin^{2}\theta_{0}=
\frac{1}{1+b^{2}}
\left(1-\frac{a^{2}(1-b^{2})\sin^{2}\theta}
{a^{2}+(1+b^{2})\cos^{2}\theta}\right)
\end{equation}
and
\begin{equation}
A=\frac{a b  \sin 2 \theta \sqrt{1+a^2+b^2}}{\left(1+b^2\right)
   \left(a^2+\left(1+b^2\right) \cos^2 \theta\right)}.
\end{equation} 

\begin{table}[t]
\begin{center}
\begin{tabular}{ccc}
\hline
\hline
Parameterization & $\sin^2{\theta_{12}}$ & $\sin^2{\theta_{23}}$ \\
\hline
\hline
$U_{(23)}$ & $1-\frac{a^2}{a^2+\left(1+b^2\right) \cos ^2\theta}$ & 
$\sin^{2}\theta_{0}+A \cos\phi$ \\
$U_{(13)}$ & $\frac{a^2}{a^2+\left(1+b^2\right) \cos ^2\theta}$ & 
$\sin^{2}\theta_{0}+A \cos\phi$   \\
$U^{(13)}$ & $\sin^{2}\theta_{0}+A \cos\phi$ & $\frac{a^2}{a^2+\left(1+b^2\right) \cos ^2\theta}$ \\
$U^{(12)}$ & $\sin^{2}\theta_{0}+A \cos\phi$ & $1-\frac{a^2}{a^2+\left(1+b^2\right) \cos ^2\theta}$\\
\hline
\end{tabular}
\end{center}
\caption{The expressions for $\theta_{12}$ and $\theta_{23}$ in terms of the
parametrizations proposed in the present work.}
\end{table}

Using various partial mixing schemes, these relations
can be used to obtain sum rules for the mixing angles and 
\textit{CP} violation by eliminating the
parameters $\theta$ and $\phi$. The sum rules relating 
$\theta_{12}$ and $\theta_{13}$ are
$\cos\theta_{13}\cos\theta_{12}=a N$ 
and $\cos\theta_{13}\sin\theta_{12}=a N$ 
for the partial mixing schemes $C_{1}$ and $C_{2}$, respectively. The sum rule for 
$\theta_{23}$ and $J$ is 
\begin{equation}
\frac{(\sin^{2}\theta_{23}-\sin^{2}\theta_{0})^{2}}{A^{2}}+
\frac{J^{2}}{B^{2}}=1
\end{equation}
for both the partial mixing schemes $C_{1}$ and $C_{2}$. 
Similarly, one can easily see that
$\cos\theta_{13}\sin\theta_{23}=a N$ 
and $\cos\theta_{13}\cos\theta_{23}=a N$ for the partial mixing schemes $R_{2}$ and $R_{3}$, respectively. For both of these mixing schemes, $\theta_{12}$ and $J$
are related as 
\begin{equation}
\frac{(\sin^{2}\theta_{12}-\sin^{2}\theta_{0})^{2}}{A^{2}}+
\frac{J^{2}}{B^{2}}=1.
\end{equation}
Hence, the relation between $\theta_{23}$ ($\theta_{12}$) and $J$ is equation 
for an ellipse centered around the point $(\sin^{2}\theta_{0},0)$ for the mixing schemes 
$C_{1}$ and $C_{2}$ ($R_{1}$ and $R_{2}$). The semiminor axis of the ellipse equals B, 
the value of J for maximal \textit{CP} violation, whereas the semimajor axis is given by 
$A$, the 
deviation of $\sin^{2}\theta_{23}$ ($\sin^{2}\theta_{12}$) from $\sin^{2}\theta_{0}$. 
A generic prediction of these relations is that \textit{CP} violation will be maximal
for a maximal $\theta_{23}$ ($\theta_{12}$) for the mixing schemes $C_{1}$
and $C_{2}$ ($R_{1}$ and $R_{2}$). 
 
The detailed phenomenology of the resulting partial mixing schemes have been studied 
extensively in the \mbox{literature \cite{modifications}}. The results of some of these 
studies can readily be obtained from the above relations and sum-rules 
by substituting respective values of the symmetry parameters $a$ and $b$ listed in 
\mbox{Table 1}. (However, we differ from many of these studies in the way we introduce
the phase $\phi$ in the mixing matrix.) 
Since all these studies presume certain values for the parameters $a$ and $b$ 
at the outset, they do not address the following two questions:
\begin{enumerate}
\item What is the allowed parameter space for the parameters $a$ and $b$?
\item How do the symmetry based values for the parameters $a$ and $b$ 
(\mbox{Table 1}) 
compare with the experimentally allowed values for these parameters?
\item Which of the partial mixing matrices are preferred by the current neutrino
mixing data?
\end{enumerate}
Here, we shall address these questions in a model independent manner as an 
important application of
the parametrizations we have presented here.

\section{The factorization of the mixing matrix}
One of the main advantages of the parametrizations proposed here is that the mixing 
matrix can be factorized in two parts:
\begin{equation}
U_{(ij)}=V_{(ij)}(a,b)R_{(ij)}(\theta,\phi)
\end{equation} 
and 
\begin{equation}
U^{(ij)}=R^{(ij)}(\theta,\phi)V^{(ij)}(a,b).
\end{equation} 
In these equations, the matrices $V_{(ij)}(a,b)$ and 
$V^{(ij)}(a,b)$ are simply the value of 
$U_{(ij)}$ and $U^{(ij)}$ for $\theta=0$ and $\phi=0$.
The complex rotations are given by $R_{(ij)}(\theta,\phi)=P(\phi)O_{(ij)}(\theta)$ and 
$R^{(ij)}(\theta,\phi)=O_{(ij)}(\theta)Q(\phi)$.
Here, the indices $\{i,j,k\}$ are cyclic permutations of $\{1,2,3\}$. 
The matrix $O_{(ij)}$ is an orthogonal $(i,j)$ rotation by angle $\theta$,
$P(\phi)=\text{diag}\{1,1,e^{i \phi}\}$ and $Q(\phi)=\text{diag}\{e^{i \phi},1,1\}$.

The factorization of the mixing matrix in the two parts $V(a,b)$ and $R(\theta,\phi)$ has
many important implications. One can associate the $V(a,b)$ part with a 
complete mixing scheme like TBM and the $R(\theta,\phi)$ part with a modification or
perturbation to that scheme. We note that the $V(a,b)$ part reduces to
the complete mixing schemes listed in \mbox{Table 1} for the respective values of the
parameters $a$ and $b$ (except for TFH1 and TFH2 where $\theta_{13}$ is
nonzero in the complete mixing scheme itself). Therefore, the perturbation 
$R(\theta,\phi)$
affects only two parameters ($\theta$ and $\phi$) from their values in the 
corresponding full mixing ($\theta=0$ and $\phi=0$) . The parameters $a$ and $b$
remain unaffected by the perturbation. In other words, the parametrizations 
proposed here parametrize not only the four experimental observables in the 
mixing matrix in terms of the four parameters $a$, $b$, $\theta$ and $\phi$), 
they can also parametrize the neutrino mass 
matrices giving rise to partial mixing schemes. In this model building context,  
the $V(a,b)$ part can come from a residual flavor symmetry and  the $R(\theta,\phi)$ 
part can result from some symmetry breaking terms. 

The parametrizations $U^{(ij)}$ have another interesting
property. The lepton mixing matrix can be written as $V_{l}^{\dagger}V_{\nu}$,
where $V_{l}$ and $V_{\nu}$ are the unitary matrices that diagonalize the charged lepton
and effective neutrino mass matrices, respectively. A comparison with 
\mbox{Eq. (14)} yields 
$V_{l}=R^{(ij)^{\dagger}}(\theta,\phi)=Q(-\phi)O_{(ij)}^{T}(\theta)$ and 
$V_{\nu}=V^{(ij)}(a,b)$. Therefore, the $V^{(ij)}(a,b)$ part can be viewed as the neutrino 
mixing coming from the neutrino mass matrix and the $R^{(ij)^{\dagger}}(\theta,\phi)$ 
part can be viewed as resulting from the charge lepton corrections.

\begin{figure*}[t]
\centering 
\centering 
\subfigure[$U_{(23)}$ ($C_{1}$)]{\includegraphics[scale=0.5]{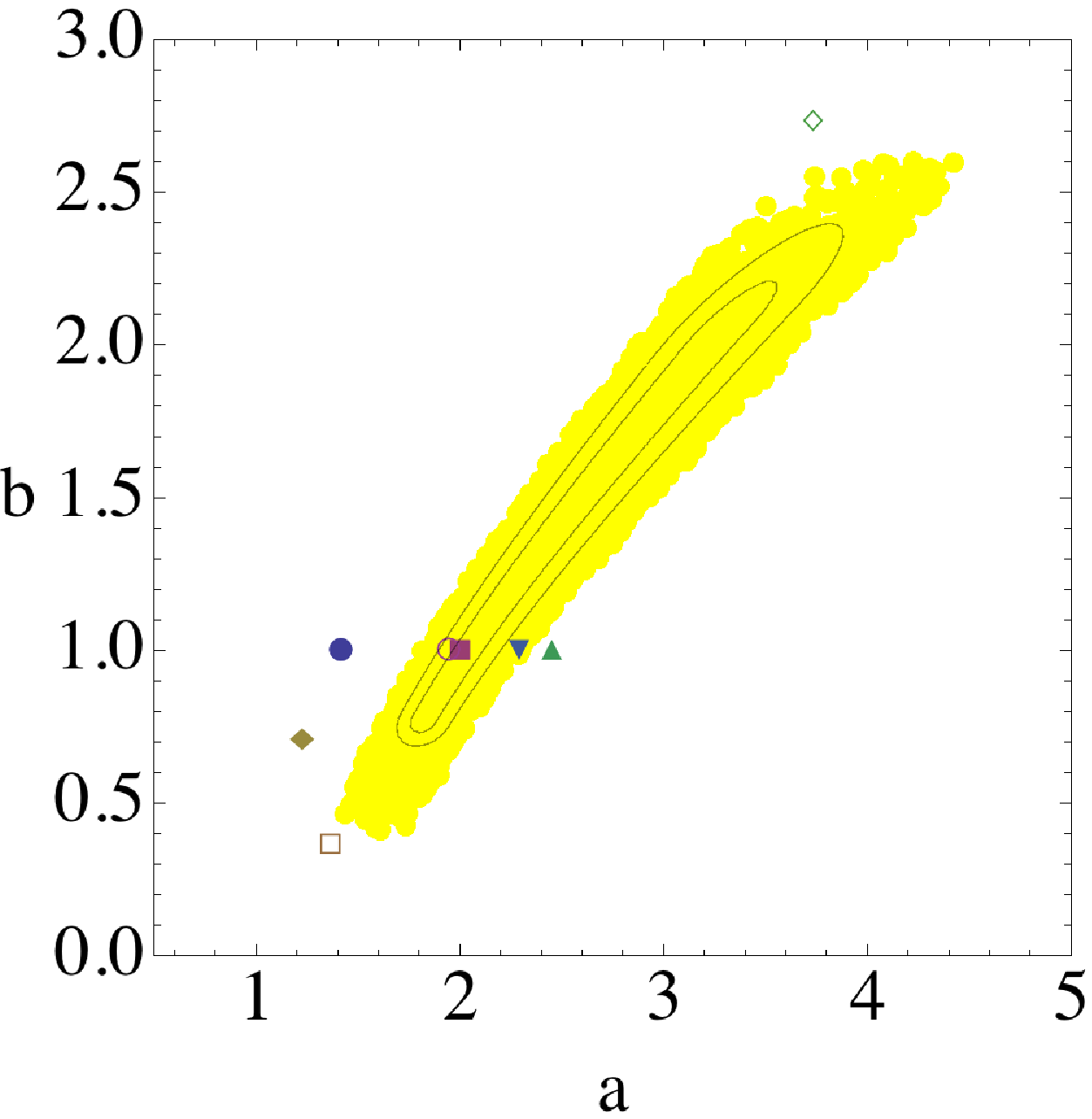}} 
\subfigure[$U_{(13)}$ ($C_{2}$)]{\includegraphics[scale=0.5]{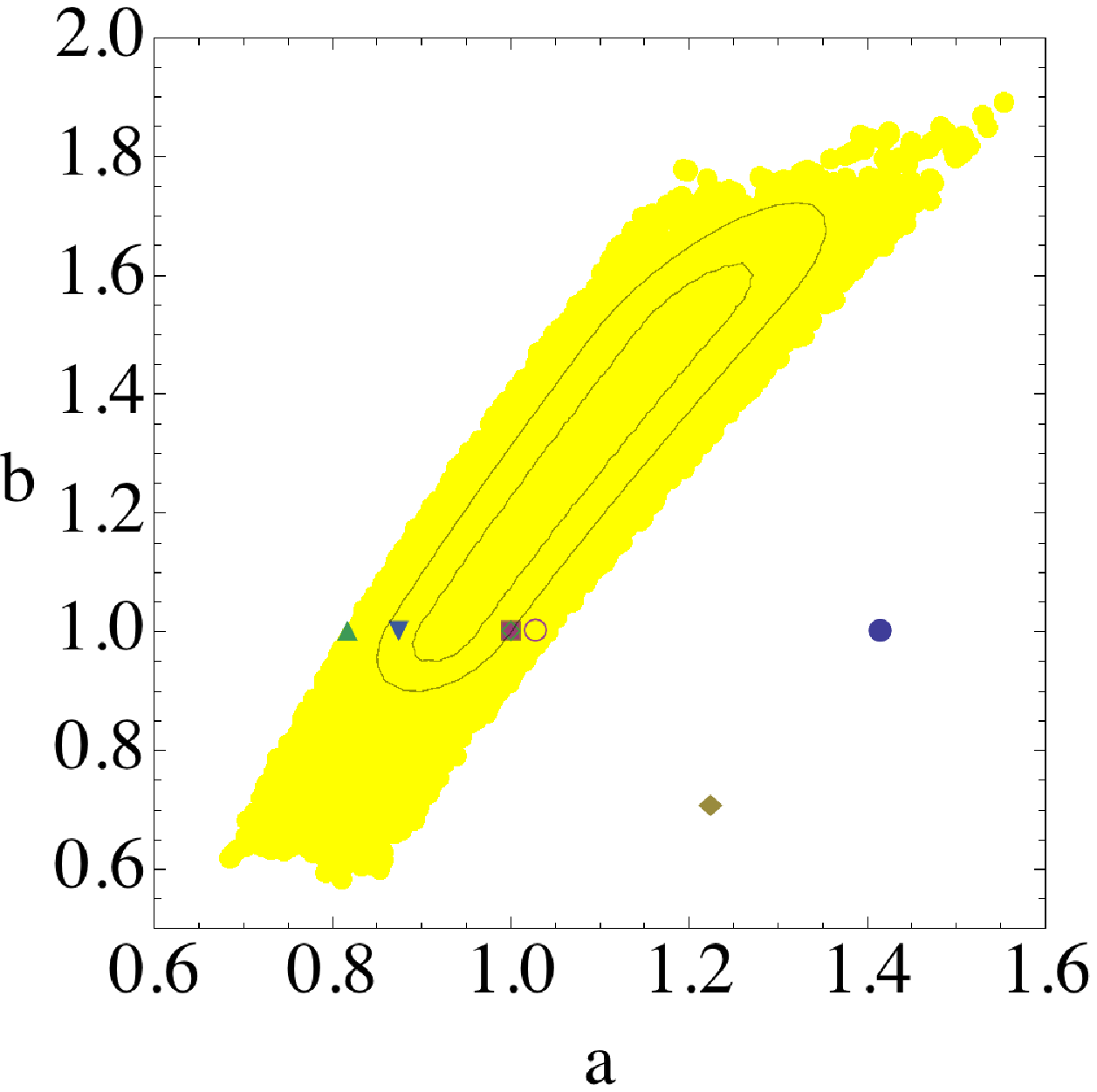}} 
\subfigure[$U^{(13)}$ ($R_{2}$)]{\includegraphics[scale=0.5]{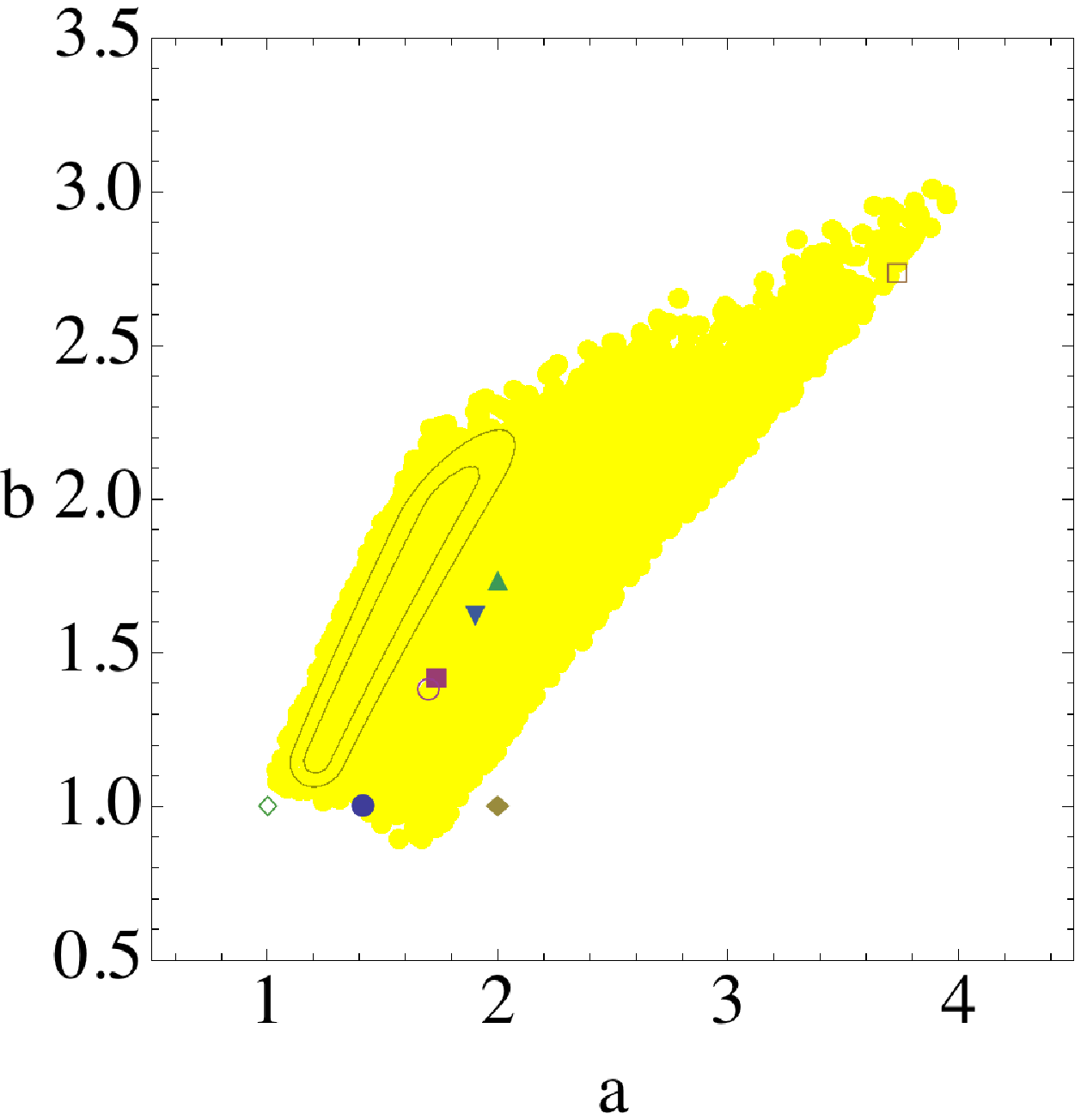}} 
\subfigure[$U^{(12)}$ ($R_{3}$)]{\includegraphics[scale=0.5]{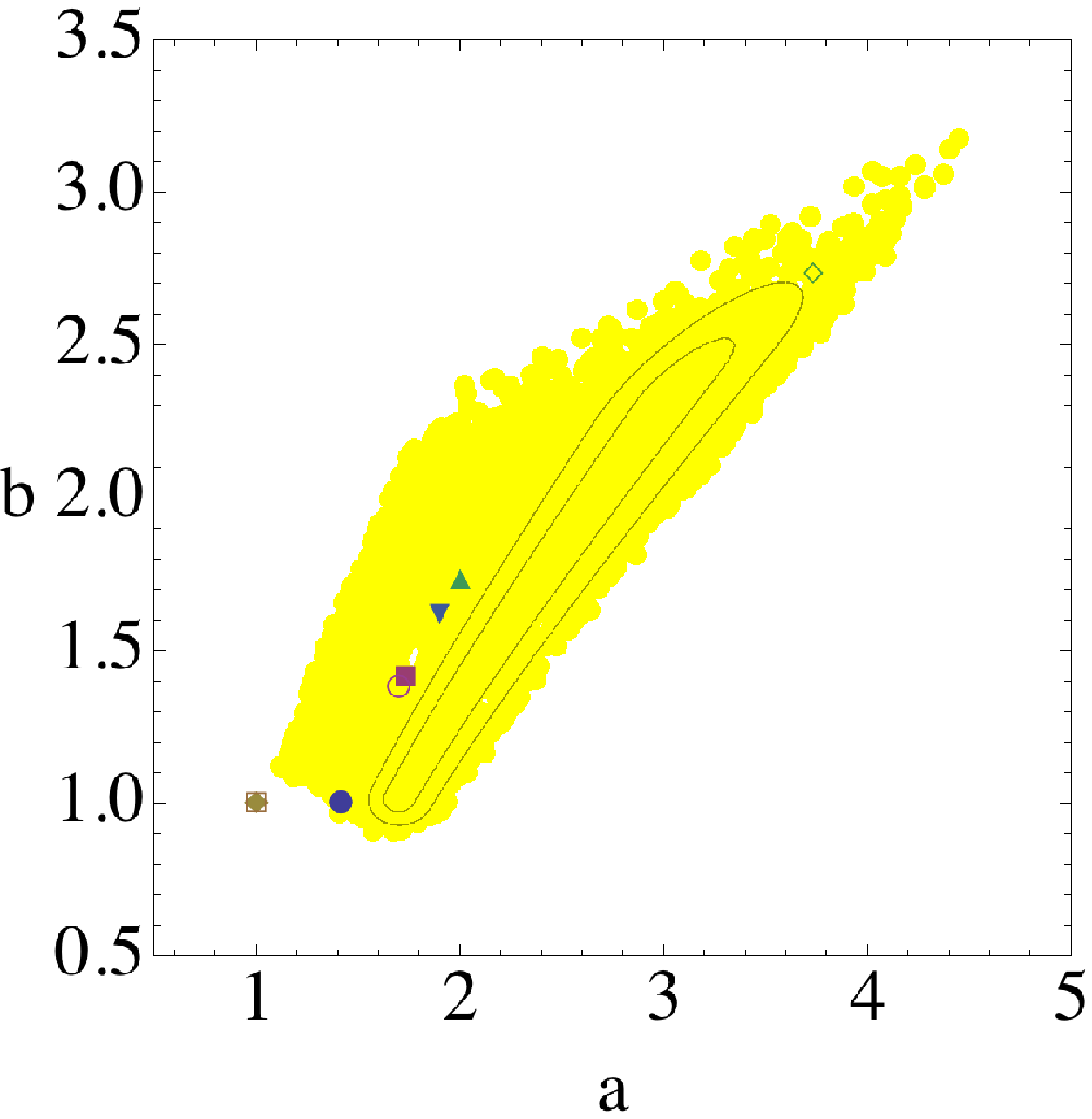}} 
\subfigure{\includegraphics[scale=0.6]{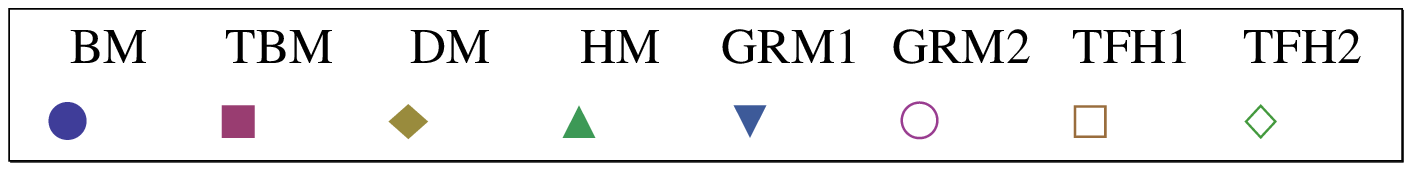}} 
\caption{(color online). The allowed parameter space for the parameters $a$ and $b$. 
The two contours
give the $1 \sigma$ and $2 \sigma$ allowed region for the parameters $a$ and $b$ 
when the other two parameters ($\theta$ and $\phi$) 
are being marginalized away. The gray (yellow) regions depict the allowed parameter space
at $3\sigma$. The symmetry values of  $a$ and $b$ for 
corresponding to various mixing patterns (\mbox{Table 1}) are also depicted for comparison. The values of $a$ and $b$ coincide for TBM, TFH1, and TFH2 for
$C_{2}$ and TFH1 and DM for $R_{3}$.}
\end{figure*}

\section{Analysis and results}

We first perform a Monte Carlo analysis of the allowed parameter space for the 
parameters $a$ and $b$ to get the preferences in the present experimental data for
various partial mixing schemes. We start with generating $n$ random 
samples for the parameters  
$a$, $b$, $\theta$ and $\phi$ with uniform 
distributions in the ranges $\{a_{min},a_{max}\}$, $\{b_{min},b_{max}\}$, 
$\{\theta_{min},\theta_{max}\}$ and $\{\phi_{min},\phi_{max}\}$.
These values are used to calculate the mixing angles $\theta_{12}$, $\theta_{23}$,
and $\theta_{13}$ from the relations given in Sec. 3.
The allowed parameter space on the plane $(a,b)$ is depicted in Fig. 1
where the three mixing angles $\theta_{12}$, $\theta_{23}$ and $\theta_{13}$
are consistent with their experimental \mbox{values \cite{data}} at 3 standard deviations for all the four parametrizations. 
The ranges $\{a_{min},a_{max}\}$, $\{b_{min},b_{max}\}$, 
$\{\theta_{min},\theta_{max}\}$ and $\{\phi_{min},\phi_{max}\}$
are decided by the following algorithm: begin with arbitrarily small ranges, say $\{0,1\}$, 
and then go on gradually expanding them untill the allowed parameter space does not
expand further. 

The main motivation of the above analysis is to obtain an understanding of the 
experimental viability of different modifications of some popular mixing schemes. 
However, it does not give us the best fit values and their experimental errors for the 
various parameters. This is accomplished by doing a $\chi^{2}$ analysis for the three mixing angles $\theta_{12}$, $\theta_{23}$ and $\theta_{13}$. The allowed regions
are depicted in Fig. 1 as contours at $1\sigma$ and $2\sigma$ C.L. The allowed
ranges for the parameters are tabulated in \mbox{Table 3}. 

\begin{table}[td]
\begin{center}
\begin{tabular}{cccc}
\hline
\hline
& $a$ & $b$ & $\theta$ \\
\hline
\\
\multirow{3}{*} {$U_{(23)}$} & 1.75-3.55 & 0.73-3.53 & 0.26-0.29 \\
& 1.69-3.89 & 0.68-2.39 & 0.24-0.31 \\
& 1.64-4.14 & 0.60-2.53 & 0.21-0.35 \\
\\
\multirow{3}{*}{$U_{(13)}$} & 0.89-1.27 & 0.91-1.61 & 0.18-0.20 \\
& 0.85-1.36 & 0.89-1.72 & 0.17-0.21 \\
& 0.67-1.57 & 0.54-1.88 & 0.15-0.23 \\
\\
\multirow{3}{*}{$U^{(13)}$} & 1.16-1.93 & 1.11-2.10 & 0.19-0.21 \\
& 1.10-2.07 & 1.09-2.22 & 0.18-0.22\\
&0.99-4.01 & 0.85-3.01 & 0.16-0.29 \\
\\
\multirow{3}{*}{$U^{(12)}$} & 1.63-3.34 & 0.96-2.51 & 0.23-0.26 \\
& 1.55-3.69 & 0.92-2.69 & 0.22-0.27 \\
& 1.07-4.47 & 0.82-3.12 & 0.16-0.31 \\
\\
\hline
\end{tabular}
\end{center}
\caption{The experimentally allowed values of the parameters $a$, $b$ and $\theta$
for the four parametrizations. The successive rows give the allowed ranges at $1$,
$2$ and $3$ $\sigma$ C.L. The phase $\phi$ can take any value in its full range 
$\{0,2\pi\}$.}
\end{table}

Finally, we superimpose the values of the parameters $a$ and $b$ listed in \mbox{Table 1} 
corresponding to various mixing schemes on the allowed parameter regions depicted in 
\mbox{Fig. 1}.  A comparison between the symmetry based values for the parameters $a$ 
and $b$ (\mbox{Table 1}) and the experimentally allowed values for these parameters can 
be made visually from \mbox{Fig. 1}. The confidence levels at which the various partial
mixing schemes are allowed or ruled out by this 
model-independent analysis are given in \mbox{Table 4}. 
We note that the partial mixing schemes constructed from all four partial mixing 
schemes constructed from DM are disallowed by the current mixing data at more than
$3\sigma$ C.L. The partial mixing matrices for BM mixing are disallowed at more than
$3\sigma$ for $C_{1}$ and $C_{2}$ and at more than $2\sigma$ C.L. for $R_{1}$
and $R_{2}$. The partial mixing schemes for HM of the types $C_{2}$, $R_{2}$ and 
$R_{3}$ are disallowed at $2\sigma$ C.L. whereas the partial mixing of type $C_{1}$ for 
HM is disallowed at more than $3\sigma$ C.L. In fact, all of the partial mixing 
patterns of type $R_{2}$ and $R_{3}$ are ruled out at more than 
$2\sigma$ C.L. because of a preference for 
$\theta_{23}<\pi/4$ in the present mixing data at $2\sigma$ C.L. \cite{data}.
The most successful mixing pattern is $C_{1}$ for TBM. This partial mixing matrix is even more 
successful than TM mixing (partial mixing of type $C_{2}$ for TBM) and have  
already been studied in the literature \cite{tm1} where its model realization has also been
discussed. Two similar partial mixing matrices of type $C_{1}$ for GRM2 and of type 
$C_{2}$ for GRM1 are also viable and should be considered for model building.

\begin{table}[td]
\begin{center}
\begin{tabular}{ccccc}
\hline
\hline
 & $C_{1}$ & $C_{2}$ & $R_{2}$ & $R_{3}$ \\
 \hline
BM & $>3\sigma$ & $>3\sigma$ & $>2\sigma$ & $>2\sigma$ \\
TBM & $\mathbf{<1\sigma}$ & $\mathbf{<2\sigma}$ & $>2\sigma$ & $>2\sigma$ \\
DM & $>3\sigma$ & $>3\sigma$ & $>3\sigma$ & $>3\sigma$ \\
HM & $>3\sigma$ & $>2\sigma$ & $>2\sigma$ & $>2\sigma$ \\
GRM1 & $>2\sigma$ & $\mathbf{<2\sigma}$ & $>2\sigma$ & $>2\sigma$ \\
GRM2 &  $\mathbf{<2\sigma}$ & $>2\sigma$ & $>2\sigma$ & $>2\sigma$ \\
TFH1 &  $>3\sigma$ & $\mathbf{<2\sigma}$ & $>2\sigma$ & $>3\sigma$ \\
TFH2 &  $>3\sigma$ & $\mathbf{<2\sigma}$ & $>3\sigma$ & $>2\sigma$ \\
\hline
\end{tabular}
\end{center}
\caption{The confidence levels by which the various partial mixing schemes
listed in \mbox{Table 1} are allowed or disallowed by the neutrino mixing data. The symbol
$>n\sigma$ means that the corresponding mixing scheme is disallowed at more than
$n\sigma$ C.L. The symbol $\mathbf{<n\sigma}$ means that the corresponding mixing matrix is allowed and the disagreement with the experimental data in less than 
$n\sigma$ C.L.}
\end{table}

\section{Conclusions}
In summary, we have presented six parametrizations for the lepton mixing matrix. 
These parametrizations are useful to describe neutrino mixing and \textit{CP} violation in 
any lepton mass model possessing a residual symmetry. The parametrizations are ideal to 
describe the partial mixing schemes that are minimal modifications of 
the complete mixing schemes. As an application of these parametrizations, we
obtain interesting sum-rules for neutrino mixing angles and \textit{CP} violation for the
partial mixing schemes in a model independent manner. These parametrizations can 
factored into two parts: $U_{0}(a,b)$ and $R(\theta,\phi)$. The $U_{0}(a,b)$ part 
can be considered as the zeroth order prediction of a flavor symmetry and the 
$R(\theta,\phi)$ part can be considered as a correction to it. We find the experimentally
allowed parameter space for the parameters $a$ and $b$. The allowed ranges of these
parameters can be interpreted as the model-independent predictions for a 
neutrino mass matrix with $Z_{2}\times Z_{2}$ symmetry. We compare these predictions
for the parameters $a$ and $b$ with their values in different partial mixing 
schemes. This model independent analysis favors the minimal modifications of TBM and 
GRM schemes over the modifications of BM, DM, and HM mixing.

It is a pleasure to thank Smarajit Triambak and Radha Raman Gautam for critical reading 
of the manuscript and helpful suggestions. This work is supported by the Department of
Science and Technology (DST), Government of India \textit{vide}  Grant No.
SR/FTP/PS-123/2011.

\begin{appendix}
\numberwithin{equation}{section}

\section{The parametrization $U_{(23)}$}

As an illustration, we will find a mixing matrix of type
\begin{equation}
U=
\left(
\begin{array}{ccc}
 a N &u_1 & v_1\\
 b N & u_2  & v_2\\
 N &  u_3  & v_3\\
\end{array}
\right)
\end{equation}
from the unitarity constraints. Here, $u_1=x_1+i y_1$, 
$v_1=x_2+i y_2$, $u_2=x_3+i y_3$ and $v_2=x_4+i y_4$,
The orthogonalization of the columns of $U$ yields
$u_3=-(a u_1+b u_2)$ and $v_3=-(a v_1+b v_2)$. Substituting
these values in \mbox{Eq. (A.1)}, we obtain
\begin{widetext}
\begin{equation}
U=
\left(
\begin{array}{ccc}
 a N & x_{1}+i y_{1} &  x_{2}+i y_{2}\\
 b N & x_{3}+i y_{3} &  x_{4}+i y_{4}\\
N & -(a x_{1}+b x_{3})- i(a y_{1}+b y_{3}) & 
-(a x_{2}+b x_{4})-i(a y_{2}+b y_{4}) \\
\end{array}
\right).
\end{equation}
Further, solving the unitarity relations $UU^{\dagger}=U^{\dagger}U=1$, we get
\begin{equation}
x_2^{2}=1-a^2 N^2-x_1^2-y_1^2-y_2^2,
\end{equation}
\begin{equation}
y_3=\frac{1}{(1+b^2)^2}\{-a b (1+b^2) y_1 +d\},
\end{equation}
\begin{equation}
x_4=\frac{a b x_2}{1+b^2}-\frac{x_1x_2+y_1y_2}{x_2^2+y_2^2}\left(x_3+\frac{a b x_1}{1+b^2}\right)-
\frac{d(x_1y_2+y_1x_2)}{(1+b^2)^2(x_2^2+y_2^2)}
\end{equation}
and
\begin{equation}
y_4=-\frac{a b y_2}{1+b^2}-\frac{x_1y_2-y_1x_2}{x_2^2+y_2^2}\left(x_3+\frac{a b x_1}{1+b^2}\right)-
\frac{d(x_1x_2+y_1y_2)}{(1+b^2)^2(x_2^2+y_2^2)},
\end{equation}
where 
\begin{equation}
d^2=(1+b^2)^2\left[(1+a^2+b^2)c^2(x_2^2+y_2^2)-\{a b x_1+(1+b^2)x_3\}^2\right].
\end{equation}
\end{widetext}
The parameters $y_1$ and $y_2$ give rise to the 
Majorana phases which can be factored out into unconstrained  neutrino masses. 
Substituting $y_{1}=y_{2}=0$, we are effectively left with two free parameters 
in the mixing matrix
\textit{viz.} $x_1$ and $x_3$, 
a fact that can be checked from simple parameter counting. These parameters can be 
further reparametrized in terms of two angles $\theta$ and $\phi$ as
$
x_1=N\sqrt{1+b^2}\cos \theta
$
and
$
\sqrt{1+b^2} x_3= \sin \theta \cos\phi-a b N \cos \theta.
$
With these redefinitions, the most general mixing matrix of type $C_{1}$ given by 
\mbox{Eq. (A.2)} becomes
\begin{equation}
U=\left(
\begin{array}{ccc}
a N & N\sqrt{1+b^2} \cos \theta &
   N\sqrt{1+b^2} \sin \theta\\
 b N & \frac{e^{i \phi} \sin \theta-a b N\cos \theta}{\sqrt{1+b^2}} &
   \frac{-c e^{i \phi} \cos \theta-a b N \sin
   \theta}{\sqrt{1+b^2}} \\
  N &\frac{-a N \cos \theta-b e^{i \phi} \sin \theta}{\sqrt{1+b^2}} &
   \frac{b e^{i \phi} \cos \theta-a N \sin
   \theta}{\sqrt{1+b^2}}
\end{array}
\right).
\end{equation}

\end{appendix}

\end{document}